\input harvmac
\sequentialequations

\def\<{}\def\>{}
\def\nofigures{\let\<=\iffalse\let\>=\fi}
\< \input pictex \>

\def\stackrel#1#2{\mathrel{\mathop{#2}\limits^{#1}}}
\def\gsim{\;\lower.5ex\hbox{$\stackrel{\lower.1ex\hbox{$>$}}{\sim}$}\;}
\hyphenation{co-ordinate co-ordinates ergo-region ergo-regions}

\Title{\baselineskip=14pt\vbox{\rightline{DAMTP R98/17}
\rightline{gr-qc/9803098}}}
{\vbox{\centerline{Rotating traversable wormholes}}}

\centerline{Edward Teo}
\bigskip
\centerline{\sl Department of Applied Mathematics and Theoretical 
Physics, University of Cambridge,}
\centerline{\sl Silver Street, Cambridge CB3 9EW,
England}
\medskip\centerline{and}
\medskip
\centerline{\sl Department of Physics, National University of 
Singapore, Singapore 119260}

\vskip 1.2in
\centerline{\bf Abstract}
\medskip
\noindent
The general form of a stationary, axially symmetric traversable 
wormhole is discussed. This provides an explicit class of rotating 
wormholes that generalize the static, spherically symmetric
ones first considered by Morris and Thorne. In agreement with 
general analyses, it is verified that such a wormhole generically 
violates the null energy condition at the throat. However, for 
suitable model wormholes, there can be classes of geodesics 
falling through it which do not encounter any 
energy-condition-violating matter. The possible presence of 
an ergoregion surrounding the throat is also noted.

\Date{PACS: 04.20.Cv, 04.20.Gz, 04.20.Jb}

\nref\MT{M.S. Morris and K.S. Thorne, Am. J. Phys. {\bf 56}, 395 
(1988).}

\nref\ER{A. Einstein and N. Rosen, Phys. Rev. {\bf 48}, 73 (1935).}

\nref\Wheeler{J.A. Wheeler, {\it Geometrodynamics\/} (Academic Press, 
New York, 1962).}

\nref\HVI{D. Hochberg and M. Visser, Phys. Rev. D {\bf 56}, 4745 
(1997); in {\it The Internal Structure of Black Holes and Spacetime 
Singularities\/}, edited by L.M. Burko and A. Ori (Institute of 
Physics Press, Bristol, 1997).}

\nref\HVII{D. Hochberg and M. Visser, `The null energy condition 
in dynamic wormholes' (gr-qc/9802048); `Dynamic wormholes, 
anti-trapped surfaces, and energy conditions' (gr-qc/9802046).}

\nref\Teo{E. Teo, `Traversable wormholes --- light at the end 
of the tunnel' (unpublished).}

\nref\MTY{M.S. Morris, K.S. Thorne, and U. Yurtsever, Phys. Rev. 
Lett. {\bf 61}, 1446 (1988).}

\nref\FN{V.P. Frolov and I.D. Novikov, Phys. Rev. D {\bf 42}, 1057 
(1990).}

\nref\Frolov{V.P. Frolov, Phys. Rev. D {\bf 43}, 3878 (1991).}

\nref\KT{S.W. Kim and K.S. Thorne, Phys. Rev. D {\bf 43}, 3929 (1991).} 

\nref\Hawking{S.W. Hawking, Phys. Rev. D {\bf 46}, 603 (1992).}

\nref\VisserII{M. Visser, Phys. Rev. D {\bf 43}, 402 (1991).}

\nref\HPS{D. Hochberg, A. Popov, and S.V. Sushkov, Phys. Rev. Lett. 
{\bf 78}, 2050 (1997).}

\nref\Visser{M. Visser, {\it Lorentzian Wormholes: From Einstein to 
Hawking\/} (AIP Press, Woodbury, N.Y., 1995).}

\nref\VisserI{M. Visser, Phys. Rev. D {\bf 39}, 3182 (1989).}

\nref\Roman{T.A. Roman, Phys. Rev. D {\bf 47}, 1370 (1993).}

\nref\KarI{S. Kar, Phys. Rev. D {\bf 49}, 862 (1994).}

\nref\KarII{S. Kar and D. Sahdev, Phys. Rev. D {\bf 53}, 722 (1996).}

\nref\SA{F. Schein and P.C. Aichelburg, Phys. Rev. Lett. {\bf 77}, 
4130 (1996).}

\nref\Papapet{A. Papapetrou, Ann. Inst. H. Poincar\'e {\bf A4}, 
83 (1966).}

\nref\CarterI{B. Carter, J. Math. Phys. {\bf 10}, 70 (1969);
Commun. Math. Phys. {\bf 17}, 223 (1970).}

\nref\Carter{B. Carter, in {\it Gravitation in Astrophysics\/},
edited by B. Carter and J.B. Hartle (Plenum Press, New York, 1987).}

\nref\Islam{J.N. Islam, {\it Rotating Fields in General Relativity\/}
(Cambridge University Press, Cambridge, 1985).}

\nref\Thorne{K.S. Thorne, in {\it General Relativity and 
Cosmology\/}, edited by R.K. Sachs (Academic Press, New York, 1971).}

\nref\HartleI{J.B. Hartle and D.H. Sharp, Astrophys. J. {\bf 147}, 
317 (1967).}

\nref\HartleII{J.B. Hartle, Astrophys. J. {\bf 150}, 1005 (1967).}

\nref\Felice{F. de Felice and C.J.S. Clarke, {\it Relativity on Curved 
Manifolds\/} (Cambridge University Press, Cambridge, 1990).}

\nref\HE{S.W. Hawking and G.F.R. Ellis, {\it The Large Scale Structure 
of Space-time\/} (Cambridge University Press, Cambridge, 1973).}

\nref\Wald{R.M. Wald, {\it General Relativity\/} (University of 
Chicago Press, Chicago, 1984).}

\nref\PenroseI{R. Penrose, Rivista del Nuovo Cimento {\bf 1}, 252 
(1969).}

\nref\PenroseII{R. Penrose and R.M. Floyd, Nature (Phys. Sci.) 
{\bf 229}, 177 (1971).}

\nref\EGJ{H. Epstein, V. Glaser, and A. Jaffe, Nuovo Cimento {\bf 36},
1016 (1965).}

\nref\AHS{P.R. Anderson, W.A. Hiscock, and D.A. Samuel, Phys. Rev. D 
{\bf 51}, 4337 (1995).}

\nref\Khatsym{V.M. Khatsymovsky, `Rotating vacuum wormhole'
(gr-qc/9803027).}

\newsec{Introduction}

The concept of a {\it traversable wormhole\/} was first put forward 
by Morris and Thorne \MT\ in 1988. Unlike those previously considered,
such as the Einstein--Rosen bridge \ER\ or the microscopic 
charge-carrying wormholes of Wheeler \Wheeler, traversable wormholes 
by definition permit the two-way travel of objects like human beings. 
Despite the dubious possibility of ever creating or finding such a 
wormhole, their study has opened up remarkably fruitful avenues of 
research. These include the fundamental properties of such wormholes 
\refs{\HVI{--}\Teo}, their use as time-machines \refs{\MTY,\FN} 
and the associated problems of causality violation 
\refs{\Frolov{--}\Hawking}, as well as the structure of quantum 
or Planck-scale wormholes \refs{\VisserII,\HPS}.

Perhaps the key feature in the analysis of Morris and Thorne \MT, is 
that they first list the conditions that a traversable wormhole must 
satisfy, and then use the Einstein equations to deduce the form of the 
matter required to maintain the wormhole. This is opposite to the 
usual procedure of postulating the matter content, and then solving the 
Einstein equations to obtain the space-time geometry (a step which is
often very difficult, if not impossible). The paradigm shift enables
a surprising amount of information to be deduced about such wormholes.

They considered a static, spherically symmetric and asymptotically flat 
space-time with the metric
\eqn\MTmetric{{\rm d}s^2=-{\rm e}^{\Phi(r)}{\rm d}t^2+\left(1
-{b(r)\over r}\right)^{-1}{\rm d}r^2+r^2\left({\rm d}\theta^2
+\sin^2\theta{\rm d}\varphi^2\right),}
where $\Phi$ and $b$ are two arbitrary functions of $r$ known as the 
redshift and shape functions respectively. The former determines 
the gravitational redshift of an infalling object, while the 
latter characterizes the shape of the wormhole as seen using an
embedding diagram; hence their names. It was shown in Ref.~\MT\ that 
for this metric to describe a wormhole, $b$ must satisfy a certain 
flare-out condition, in which case \MTmetric\ describes two identical 
asymptotic universes joined together at the `throat' $r=b$. The 
condition that the wormhole be traversable, in particular, means 
that there are no event horizons or curvature singularities. This 
translates to the requirement that $\Phi$ be finite everywhere.

Morris and Thorne \MT\ then went on to prove that the metric \MTmetric,
together with the wormhole-shaping and traversality conditions on $b$ 
and $\Phi$, imply that the corresponding stress-energy tensor 
necessarily violates the null (and therefore also the weak) energy 
condition \Visser. They called this form of matter `exotic', an 
acknowledgement of the fact that there is an astronomical, perhaps 
impossible, price to be paid for interstellar travel using wormholes.

This has not prevented some authors from studying other classes of
traversable wormholes, in the hope of minimizing the violation of 
the energy conditions. It was realized early on that by giving up 
spherical symmetry, it is possible to move the exotic matter 
around in space so that some observers falling through the
wormhole would not encounter it \VisserI. This was demonstrated
by cutting out holes in Minkowski space, and joining them up with 
a thin (delta-function) layer of matter.

Another natural generalization of the work of Morris and Thorne 
is to include time dependence. Perhaps the simplest way is to include 
a time-dependent conformal factor in the metric \MTmetric,  while 
preserving spherical symmetry \refs{\Roman{--}\KarII}:
\eqn\nameless{{\rm d}s^2=\Omega^2(t)\left[-{\rm e}^{\Phi(r)}{\rm d}t^2
+\left(1-{b(r)\over r}\right)^{-1}{\rm d}r^2+r^2\left({\rm d}\theta^2
+\sin^2\theta{\rm d}\varphi^2\right)\right].}
When $\Omega(t)$ is increasing, this metric represents a conformally 
expanding Morris--Thorne wormhole. It has been proposed that this 
might describe a wormhole being `pulled out' of the space-time 
foam to a macroscopic size during the inflationary epoch \Roman. 
If the wormhole were expanding fast enough, it appears possible 
to avoid any violation of the energy conditions. But as explained 
in Ref.~\KarII, this is because any observer travelling through 
the wormhole would see its `radius' increasing all the way, and 
thus it would not qualify as a wormhole in the usual sense.

In this paper, I shall construct the stationary and axially symmetric 
generalization of the Morris--Thorne wormhole \MTmetric. This would 
physically describe rotating wormholes. There are a few reasons 
why considering this case is important: it is perhaps the most general 
extension of the Morris--Thorne wormhole that one can fruitfully 
consider, short of a system with no space-time symmetries (which would 
be quite impractical to analyse with present techniques). This could 
then be used to derive or model explicit wormhole solutions, both 
classical \SA\ and semi-classical \HPS, that are of interest in 
various contexts. Of course, another reason is that if an arbitrarily 
advanced civilization \MTY\ were to create a traversable wormhole, 
it would in all likelihood be aspherical or rotating. This would enable 
would-be interstellar travelers to avoid passing through and 
interacting with the exotic matter needed to maintain the wormhole.

The philosophy adopted in this paper will be the same as that of 
Ref.~\MT. I shall begin by writing down the most general metric 
respecting the above-mentioned symmetries, and examine the conditions 
under which it would describe a traversable wormhole. It is then 
explicitly shown that such a wormhole would, unfortunately but 
inevitably, violate the null energy condition for a class of null
vectors at its throat. Finally, a concrete example of a rotating 
wormhole is used to demonstrate the avoidance of the exotic matter 
by certain observers, and the possible presence of an ergoregion 
surrounding the throat.

\newsec{Canonical form of the metric}

The space-times that we are interested in will be stationary and axially 
symmetric. The former means the space-time possesses a time-like Killing 
vector field $\xi^a\equiv(\partial/\partial t)^a$ generating invariant 
time translations, while the latter means it has a space-like Killing 
vector field $\psi^a\equiv(\partial/\partial\varphi)^a$ generating 
invariant rotations with respect to the angular coordinate $\varphi$. 
It is well-known, from the work of Papapetrou and Carter, that the 
most general stationary and axisymmetric metric can be written as 
\refs{\Papapet{--}\Carter}
\eqn\pmetric{{\rm d}s^2=g_{tt}{\rm d}t^2+2g_{t\varphi}{\rm d}t
{\rm d}\varphi+g_{\varphi\varphi}{\rm d}\varphi^2+g_{ij}{\rm d}x^i
{\rm d}x^j,}
where the indices $i,j=1,2$ run over the two remaining coordinates.
This metric is uniquely determined up to coordinate transformations 
of $(x^1,x^2)$. Such transformations can be used to adapt the metric 
for specific problems. For example, the choice $g_{11}=g_{22}$ and 
$g_{12}=0$ turns it into a form which enables the Einstein equations 
to be simplified considerably (see, e.g., Ref.~\Islam).

We shall, however, use this freedom to cast the metric \pmetric\ 
into spherical polar coordinates by setting 
$g_{22}=g_{\varphi\varphi}/\sin^2x^2$ \Thorne:
\eqn\metric{{\rm d}s^2=-N^2{\rm d}t^2+{\rm e}^\mu{\rm d}r^2+r^2K^2
\left[{\rm d}\theta^2+\sin^2\theta({\rm d}\varphi-\omega{\rm d}t)^2
\right],}
where the four gravitational potentials $N$, $\mu$, $K$ and $\omega$ 
depend on $(x^1,x^2)\equiv(r,\theta)$ only. This form of the metric 
has the distinct advantage of making the physics transparent. (It was 
first used by Hartle \refs{\HartleI,\HartleII} in the study of 
relativistic rotating stars.) The quantity $\omega(r,\theta)$ is the 
angular velocity ${\rm d}\varphi/{\rm d}t$ acquired by a particle 
that falls freely from infinity to the point $(r,\theta)$, and which 
gives rise to the well-known dragging of inertial frames or 
Lense--Thirring effect in general relativity. $K(r,\theta)$ is a 
positive, nondecreasing function of $r$ that determines the `proper 
radial distance' $R$ measured at $(r,\theta)$ from the origin:
\eqn\properR{R\equiv rK\,,\qquad R_r>0\,.}
($R_r\equiv\partial R/\partial r$, etc.) Notice that $2\pi R\sin\theta$ 
can be geometrically interpreted as the proper circumference of the
circle located at coordinate values $(r,\theta)$, with $\varphi$ 
ranging from 0 to $2\pi$.

The discriminant \Carter\ of the metric \metric\ is
\eqn\discri{D^2\equiv-g_{tt}g_{\varphi\varphi}+g_{t\varphi}{}^2
=(NKr\sin\theta)^2,}
which implies that an event horizon appears whenever $N=0$ (see, e.g., 
Sec.~7.10 of Ref.~\Felice). To ensure that the metric is nonsingular 
on the rotation axis $\theta=0,\pi$, the usual regularity conditions 
on $N$, $\mu$ and $K$ have to be imposed. Essentially, this means 
their $\theta$ derivatives have to vanish on the rotation axis.

To turn the stationary, axisymmetric metric \metric\ into a form 
suitable for describing a traversable wormhole, we write
\eqn\nameless{\mu(r,\theta)=-\ln\left(1-{b(r,\theta)\over r}\right),}
in terms of the new function $b(r,\theta)$. It clearly reduces to 
the Morris--Thorne canonical metric \MTmetric\ in the limit of 
zero rotation and spherical symmetry:
\eqn\nameless{N(r,\theta)\rightarrow{\rm e}^{\Phi(r)},\qquad 
b(r,\theta)\rightarrow b(r)\,,\qquad
K(r,\theta)\rightarrow1\,,\qquad \omega(r,\theta)\rightarrow0\,.}
In analogy with that case, we shall only consider the coordinate 
range $r\geq b$ and identify the apparent singularity at 
$r=b\geq0$ with the throat of the wormhole. $N$, $K$ and $\omega$ 
are assumed to be otherwise well-behaved at the throat. Now, it 
is readily checked that the scalar curvature of the space-time 
\metric\ has the following terms of order $(r-b)^{-2}$:
\eqn\nameless{-{1\over(rK)^2}\left(\mu_{\theta\theta}
+{1\over2}\mu_\theta{}^2\right)
=-{3\over2}{1\over(rK)^2}{b_\theta{}^2\over(r-b)^2}\,.}
If the throat is to be free of curvature singularities, $b_\theta$ has 
to vanish there. Hence, the throat is located at some constant value of 
$r$. (Note that this does not mean the throat is spherically symmetric, 
since the proper radial distance $R$ in general still has a $\theta$ 
dependence coming from $K$.) 

Now, in order for the geometry \metric\ to have the shape of a wormhole, 
$b$ must satisfy a so-called flare-out condition when $r=b$. This 
can be seen by embedding it in a higher-dimensional space, following 
Ref.~\MT. For constant $t$ and $\theta$, \metric\ becomes
\eqnn\nameless
$$ \eqalignno{{\rm d}s^2&=\left(1-{b(r)\over r}\right)^{-1}{\rm d}r^2
+r^2K^2\sin^2\theta{\rm d}\varphi^2\cr&=\left(1-{\beta(\rho)\over\rho}
\right)^{-1}{\rm d}\rho^2+\rho^2{\rm d}\varphi^2,&\nameless} $$
where $\rho\equiv R\sin\theta$, and $\beta$ is defined correspondingly
in terms of $b$. The throat is at $\rho=\beta$. We shall embed this 
two-surface in an (unphysical) three-dimensional Euclidean space, 
which has the metric
\eqn\nameless{{\rm d}\tilde s^2={\rm d}z^2+{\rm d}\rho^2
+\rho^2{\rm d}\varphi^2,}
in cylindrical coordinates. This surface is then described by the 
function $z=z(\rho)$, which satisfies
\eqn\embed{{{\rm d}z\over{\rm d}\rho}=\pm\left({\rho\over\beta(\rho)}
-1\right)^{-1/2}.}
That it has the characteristic shape of a wormhole, as illustrated 
in Figs.~1 and 2 of Ref.~\MT, means the flare-out condition 
${\rm d}^2\rho/{\rm d}z^2>0$ must be satisfied at the throat. But, 
we have
\eqn\nameless{{{\rm d}^2\rho\over{\rm d}z^2}={{\rm d}\rho\over{\rm d}r}
{{\rm d}^2r\over{\rm d}z^2}\,,}
when $r=b$. Since ${\rm d}\rho/{\rm d}r$ is positive by \properR, 
the flare-out condition is equivalent to
\eqn\flare{{{\rm d}^2r\over{\rm d}z^2}={b-b_rr\over2b^2}>0\,,}
at the throat. This is precisely the same condition as in the 
Morris--Thorne wormhole \MT.

Hence, the form of $b(r,\theta)$ in $g_{11}$ is very similar to
its counterpart in the Morris--Thorne wormhole \MTmetric. Recall 
that it is possible to define a new radial coordinate $l$, in that 
case, by 
\eqn\elle{{{\rm d}l\over{\rm d}r}\equiv\pm\left(1-{b\over r}\right)
^{-1/2},}
which is well-behaved across the wormhole throat. In the present 
case, it is also possible to make this change of coordinate in the 
immediate vicinity of the throat, where $b_\theta=0$ as we observed 
above. The metric \metric\ then becomes
\eqn\lmetric{{\rm d}s^2=-N^2(l,\theta){\rm d}t^2+{\rm d}l^2+r^2(l)
K^2(l,\theta)\left[{\rm d}\theta^2+\sin^2\theta({\rm d}\varphi
-\omega(l,\theta){\rm d}t)^2\right],}
to first order in $r-r_0$, where $r_0$ is the location of the 
throat. This metric smoothly connects the two asymptotic regions of 
the space-time across the throat, unlike \metric\ which is singular
there. Of course, if $b=b(r)$ is independent of $\theta$, then 
$l$ defined by \elle\ is valid everywhere and takes the range 
$(-\infty,\infty)$. The metric \lmetric\ then covers the entire 
space-time. Without loss of generality, we may assume the wormhole 
throat is at $l=0$, so that $l$ is positive on one side of the 
throat and negative on the other. The asymptotic regions are 
at $l=\pm\infty$.

Although it is implicit in the form of the above metric that the
two regions connected by the wormhole are isometric, it is 
possible to generalize it to the case when they are not. The 
static, spherically symmetric case was discussed in Sec.~11.2 of
Ref.~\Visser; and the same methods apply here. One simply introduces 
different gravitational potentials, $N_\pm$, $b_\pm$, $K_\pm$, 
$\omega_\pm$, for each of the regions labeled by $\pm$. However, 
they have to match up appropriately at the throat, to ensure that 
the metric components (in terms of $l$) are continuous and 
differentiable there.

We also require that the metric \metric\ be asymptotically flat, 
in which case
\eqn\nameless{N\rightarrow1,\qquad {b\over r}\rightarrow0\,,\qquad
K\rightarrow1\,,\qquad \omega\rightarrow0\,,}
as $r\rightarrow\infty$. Thus, $r$ is asymptotically the proper
radial distance. In particular, if
\eqn\nameless{\omega={2a\over r^3}+{\rm O}\left({1\over r^4}\right),}
then by changing to Cartesian coordinates, it can be checked that 
$a$ is the total angular momentum of the wormhole. Its mass and 
charge, if any, can also be deduced in the usual manner \Visser.

To summarize, the canonical metric for a stationary, axisymmetric 
traversable wormhole can be written as
\eqn\wormhole{{\rm d}s^2=-N^2{\rm d}t^2+\left(1-{b\over r}
\right)^{-1}{\rm d}r^2+r^2K^2\left[{\rm d}\theta^2+\sin^2\theta
({\rm d}\varphi-\omega{\rm d}t)^2\right],}
where $N$, $b$, $K$ and $\omega$ are functions of $r$, and of 
$\theta$ such that it is regular on the symmetry axis $\theta=0,\pi$.
It describes two identical, asymptotically flat regions joined 
together at the throat $r=b>0$. $N$ is the analog of the redshift 
function in \MTmetric, and it has to be finite and nonzero to ensure 
that there are no event horizons or curvature singularities. $b$ is 
the shape function which satisfies $b\leq r$. At the throat itself, 
it has to be independent of $\theta$, i.e., $b_\theta=0$, and obey 
the flare-out condition $b_r<1$. $K$ determines the proper radial 
distance as in \properR, while $\omega$ governs the angular 
velocity of the wormhole.

The space-time \wormhole\ will, in general, have nonvanishing 
stress-energy tensor components $T_{tt}$, $T_{t\varphi}$ and 
$T_{\varphi\varphi}$, as well as $T_{ij}$. They have the usual 
physical interpretations; in particular, $T_{t\varphi}$ 
characterizes the rotation of the matter distribution. For a 
static, spherically symmetric wormhole, it turns out that the 
matter required to support it must have a radial tension at the 
throat exceeding its mass-energy density \MT. This signifies a 
breakdown of the celebrated energy conditions of general 
relativity \HE. As we shall now see, the latter is also 
unavoidable for the class of wormholes described by \wormhole.

\newsec{Violation of the null energy condition}

It can be shown, under very general conditions, that a traversable 
wormhole violates the averaged null energy condition in the region
of the throat \refs{\HVII,\Teo}, by using the Raychaudhuri equation 
\HE\ together with the fact that a wormhole throat by definition
defocuses light rays. However, it is often useful to carry out a 
specific analysis, as in the present case. In this section, I shall 
explicitly show that the null energy condition \refs{\HE,\Visser}
\eqn\nameless{R_{ab}k^ak^b\geq0\,,}
where $R_{ab}$ is the Ricci tensor of the space-time \wormhole,
is violated by a class of null vectors $k^a$ at the throat. This 
would allow us to determine the precise location of the violation,
and identify the gravitational potentials responsible for it.

Recall that for the Morris--Thorne wormhole \MTmetric, the null 
vectors in question are radial ones, which can be taken to be 
$k^a=(\sqrt{-g^{tt}},\pm\sqrt{g^{11}},0,0)$ \refs{\MT,\Visser}. 
This choice is also possible for the stationary, axisymmetric case
\wormhole, provided $g_{tt}<0$ everywhere. But the latter is not 
always true in a rotating system, as we shall see in the following 
section. A more natural choice turns out to be the following null 
vector:
\eqn\nameless{k^a=\left({1\over N},-{\rm e}^{-\mu/2},0,{\omega\over N}
\right),}
where we may take $N$ to be positive, without loss of generality.
In the asymptotic region, $k^a=(1,-1,0,0)$, so it represents the 
4-velocity of a (null) particle that is directed radially inwards. 
But it acquires a nonzero angular velocity $\omega$ as the throat 
is approached, due to the dragging of inertial frames \Thorne\ 
by the wormhole. We have, at the throat itself,
\eqn\NEC{R_{ab}k^ak^b={\rm e}^{-\mu}\mu_r{(rK)_r\over rK}
-{\omega_\theta{}^2\sin^2\theta\over2N^2}-{1\over4}{\mu_\theta{}^2
\over(rK)^2}-{1\over2}{(\mu_\theta\sin\theta)_\theta\over(rK)^2
\sin\theta}+{(N_\theta\sin\theta)_\theta\over(rK)^2N\sin\theta}\,.}
But by the flare-out condition \flare,
\eqn\nameless{{\rm e}^{-\mu}\mu_r={1\over r^2}(b_rr-b)<0\,.}
Furthermore, $R=rK$ is a monotonically increasing function of 
$r$. Thus, the first term on the right-hand side of \NEC\ is 
negative-definite. The second term is also manifestly 
nonpositive.

It remains to show that the sum of the remaining three terms is 
negative at some point $\theta\in[0,\pi]$ on the throat (excluding 
the trivial case when they are identically zero). We first rewrite 
them as
\eqn\fterms{\left(f_1{}^2-f_2{}^2\right)
+{(f\sin\theta)_\theta\over\sin\theta}\,,}
multiplied by an overall positive factor of $(rK)^{-2}$, where
\eqn\nameless{f_1\equiv(\ln N)_\theta\,,\qquad f_2\equiv\hbox{$1\over2$}
\mu_\theta\,,\qquad f\equiv f_1-f_2\,.}
Note that the regularity conditions on $N$ and $\mu$ imply that
$f$ vanishes at $\theta=0$ and $\pi$. Now, suppose $f<0$ at some point 
$\theta\in(0,\pi)$. By continuity, there is an interval in $(0,\pi)$ 
such that $f\sin\theta<0$ and $(f\sin\theta)_\theta<0$. The former 
means that $f_1{}^2<f_2{}^2$, and hence \fterms\ is negative in this 
interval. On the other hand, suppose $f\geq0$ everywhere in $[0,\pi]$. 
Then, as $\theta\rightarrow\pi$, we have $f\rightarrow0$ and 
$(f\sin\theta)_\theta\rightarrow0^-$. The former means that the first
term of \fterms\ vanishes in this limit, while the latter implies
that the second term of \fterms\ is negative in this limit.
Hence, we have proved that the right-hand side of \NEC\ is negative 
at some point on the throat, and there is consequently a violation of 
the null energy condition there.

A similar argument shows that the sum of the last three terms on the 
right-hand side of \NEC\ is positive at some point in the interval 
$(0,\pi)$. By choosing $N$ and $\mu$ appropriately, it is possible 
to render $R_{ab}k^ak^b$ positive at this point. Thus, the exotic 
matter supporting the wormhole can be moved around the throat, so 
that some class of infalling observers would not encounter it; 
an example is discussed in the following section. This is to be 
contrasted with the static, spherically symmetric case, where all 
such observers will experience a violation of the energy condition. 
But the key point in the above result is that one can never avoid 
the use of exotic matter altogether.

\newsec{Example of a rotating wormhole}

In the absence of any further physical input, it is possible to 
construct an infinity of wormholes of the form \wormhole, limited 
only by one's imagination. Morris and Thorne \MT\ have justifiably
argued that the use of exotic matter should be minimized. This is 
readily achieved by confining the exotic matter to a small region 
around the throat, and surrounding it with ordinary matter. They have 
also put constraints on the wormholes to make them suitable for human 
travel --- the so-called engineering considerations. However, we shall 
not be overly concerned with either of these issues. The purpose here 
is to illustrate, by means of an explicit example, some general 
features of a rotating wormhole, rather than to construct a 
semi-realistic wormhole. 

We shall take the gravitational potentials of \wormhole\ to be
\eqn\nameless{N=K=1+{(4a\cos\theta)^2\over r}\,, \qquad b=1\,,
\qquad \omega={2a\over r^3}\,. }
Perhaps the first point to note is that $a$ is the angular momentum
of the resulting wormhole. When it vanishes, the metric reduces to 
the spherically symmetric, zero-tidal-force Schwarzschild wormhole 
of Ref.~\MT. It is so-called because the embedded surface for the 
equatorial plane of the wormhole is given by (valid even for 
nonzero $a$, as can be checked using \embed)
\eqn\nameless{z(r)=\pm2\sqrt{r-1}\,,}
which is identical to that obtained by embedding the Schwarzschild
black hole space-time appropriately. The throat at $r=1$ has proper 
radius $R=1+(4a\cos\theta)^2$. Thus, it has a dumbbell-like shape as 
in Fig.~1, with minimum radius at the equator. Also note that since
$b$ is independent of $\theta$, we may introduce the new radial 
coordinate
\eqn\nameless{l=\pm\left[\sqrt{r(r-1)}+\ln\left(\sqrt{r}+\sqrt{r-1}
\right)\right],}
satisfying \elle, and the resulting metric given by \lmetric\ is 
well-behaved across the throat.

Let us consider geodesic motion in this space-time, which for 
simplicity, will be restricted to the equatorial plane $\theta=\pi/2$.
The 4-velocity vector of the geodesic is then
\eqn\nameless{k^a\equiv\dot x^a=(\dot t,\dot r,0,\dot\varphi)\,,}
where an overdot denotes derivative with respect to the affine 
parameter along the geodesic. The time-like Killing vector field 
$\xi^a$ and axial Killing vector field $\psi^a$ respectively yield a 
conserved energy $E$, and angular momentum $L$, per unit rest mass 
for the geodesic (see, e.g., Ref.~\Wald):
\eqn\consv{\eqalign{E&=-g_{ab}\xi^ak^b=-g_{tt}\dot t-g_{t\varphi}
\dot\varphi\,,\cr L&=g_{ab}\psi^ak^b=g_{t\varphi}\dot t+g_{\varphi
\varphi}\dot\varphi\,.}}
Furthermore, we have
\eqn\geod{g_{ab}k^ak^b=-\kappa\,,}
where $\kappa=0$ for null geodesics and $\kappa=1$ for time-like 
ones.

The pair of equations \consv\ can be solved to obtain
\eqn\nameless{\dot t={1\over D^2}(g_{\varphi\varphi}E+g_{t\varphi}L)\,,
\qquad \dot\varphi=-{1\over D^2}(g_{t\varphi}E+g_{tt}L)\,,}
where $D^2$ is given in \discri. Substituting this result into \geod\ 
yields an expression for $\dot r$:
\eqn\nameless{\dot r^2=g^{11}\left[{1\over D^2}(g_{\varphi\varphi}
E^2+2g_{t\varphi}EL+g_{tt}L^2)-\kappa\right].}
Thus, a geodesic freely falling towards the wormhole without any 
angular momentum would have 4-velocity
\eqn\tangent{k^a=\left({E\over N^2},-\sqrt{{\rm e}^{-\mu}
\left({E^2\over N^2}-\kappa\right)},0,{\omega E\over N^2}\right).}
When $a=1/4$ say, it can be verified that $R_{ab}k^ak^b=E^2/r^3$ 
for such geodesics which are null. This quantity is clearly 
positive. Furthermore, for time-like geodesics, we have
\eqn\wec{G_{ab}k^ak^b={1\over r^3}(E^2-3)+{9\over16}\left(
{1\over r^7}-{1\over r^6}\right),}
where $G_{ab}$ is the Einstein tensor of the space-time. When these
geodesics have sufficiently high energy $E\gsim1.75$, \wec\ is 
positive everywhere along the path. Hence, these time-like and null 
geodesics are able to traverse the wormhole without encountering 
any exotic matter having negative energy density $T_{ab}k^ak^b$. 
(After the throat is crossed, the minus sign of the $k^1$ component 
of \tangent\ becomes a plus sign.)

%
%
\< \catcode`!=11

\def\arrowhead <#1> [#2,#3]{%
  \!ifnextchar<{\!arrow{#1}{#2}{#3}}{\!arrowhead{#1}{#2}{#3}<\!zpt,\!zpt> }}

\def\!arrowhead#1#2#3<#4,#5> from #6 #7 to #8 #9 {%
%
  \!xloc=\!M{#8}\!xunit   
  \!yloc=\!M{#9}\!yunit
  \!dxpos=\!xloc  \!dimenA=\!M{#6}\!xunit  \advance \!dxpos -\!dimenA
  \!dypos=\!yloc  \!dimenA=\!M{#7}\!yunit  \advance \!dypos -\!dimenA
  \let\!MAH=\!M
  \!setdimenmode
  \!xshift=#4\relax  \!yshift=#5\relax
  \!reverserotateonly\!xshift\!yshift
  \advance\!xshift\!xloc  \advance\!yshift\!yloc
%
%
  \!Pythag\!dxpos\!dypos\!arclength
  \!divide\!dxpos\!arclength\!dxpos  
  \!dxpos=32\!dxpos  \!removept\!dxpos\!!cos
  \!divide\!dypos\!arclength\!dypos  
  \!dypos=32\!dypos  \!removept\!dypos\!!sin
%
  \!halfhead{#1}{#2}{#3}
  \!halfhead{#1}{-#2}{-#3}
  \let\!M=\!MAH
  \ignorespaces}

\catcode`!=12 \>
%
%

%
%
\topinsert
\vbox\bgroup
  \vskip5mm
  \hbox to \hsize\bgroup\hss
\<  \beginpicture
      \ninepoint
      \baselineskip=10pt

      \setcoordinatesystem units <1.0mm,1.0mm>
      \setplotarea x from -55 to 55, y from -30 to 30

      \arrow <2mm> [.2,.67] from 0 -37 to 0 43

      \ellipticalarc axes ratio 4:1 180 degrees from -5 38
        center at 0 38
      \ellipticalarc axes ratio 4:1 -45 degrees from -5 38 
        center at 0 38
      \ellipticalarc axes ratio 4:1 45 degrees from 5 38
        center at 0 38
      \arrowhead <1.2mm> [.2,.67] from 5 38.36 to 3.5 38.9

      \setquadratic
      \plot 0 30  4.62 29.15  8.68 26.72  11.75 23.06  13.57 18.68 
        14.14 14.14  13.68 9.94  12.58 6.41  11.32 3.68  10.36 1.64
        10.0 0  10.36 -1.64  11.32 -3.68  12.58 -6.41  13.68 -9.94
        14.14 -14.14  13.57 -18.68  11.75 -23.06  8.68 -26.72
        4.62 -29.15  0 -30  -4.62 -29.15  -8.68 -26.72
        -11.75 -23.06  -13.57 -18.68  -14.14 -14.14  -13.68 -9.94
        -12.58 -6.41  -11.32 -3.68  -10.36 -1.64  -10.0 0 
        -10.36 1.64  -11.32 3.68  -12.58 6.41  -13.68 9.94
        -14.14 14.14  -13.57 18.68  -11.75 23.06  -8.68 26.72
        -4.62 29.15  0 30 /

      \plot 14.14 14.14  13.99 9.94  13.62 6.41  13.20 3.68  12.88 1.64
        12.76 0  12.88 -1.64  13.20 -3.68  13.62 -6.41  13.99 -9.94  
        14.14 -14.14 /

      \plot -14.14 14.14  -13.99 9.94  -13.62 6.41  -13.20 3.68  
        -12.88 1.64  -12.76 0  -12.88 -1.64  -13.20 -3.68  -13.62 -6.41  
        -13.99 -9.94  -14.14 -14.14 /

      \setshadegrid span <1.5pt>
      \hshade -14.14 14.14 14.14  -9.94 13.68 13.99  -6.41 12.58 13.62
        -3.68 11.32 13.20  -1.64 10.36 12.88  0 10.0 12.76
        1.64 10.36 12.88  3.68 11.32 13.20  6.41 12.58 13.62
        9.94 13.68 13.99  14.14 14.14 14.14 /
        
      \hshade -14.14 -14.14 -14.14  -9.94 -13.99 -13.68  
        -6.41 -13.62 -12.58  -3.68 -13.20 -11.32  -1.64 -12.88 -10.36  
        0 -12.76 -10.0  1.64 -12.88 -10.36  3.68 -13.20 -11.32  
        6.41 -13.62 -12.58  9.94 -13.99 -13.68  14.14 -14.14 -14.14 /
        
      \put {$\varphi$} at -7 38
      \put {$\theta=0$} [l] at 2 32
      \put {$\theta=\pi/2$} [l] at 14.76 0
      \put {$\theta=\pi$} [l] at 2 -32

      \arrowhead <2mm> [.2,.67] from -10 28.1 to -8.68 26.72 
      \plot -15.68 30  -11.68 29  -8.68 26.72  /
      \put {throat} [r] at -16.68 30

      \arrowhead <2mm> [.2,.67] from -15 -4.3 to -11.5 -1.5 
      \plot -18.5 -5  -14.5 -3.5  -11.5 -1.5  /
      \put {ergoregion} [r] at -19.5 -5
    \endpicture \>
  \hss\egroup
  \vskip10mm
  \vbox{\ninepoint\leftskip0.5in\rightskip0.5in
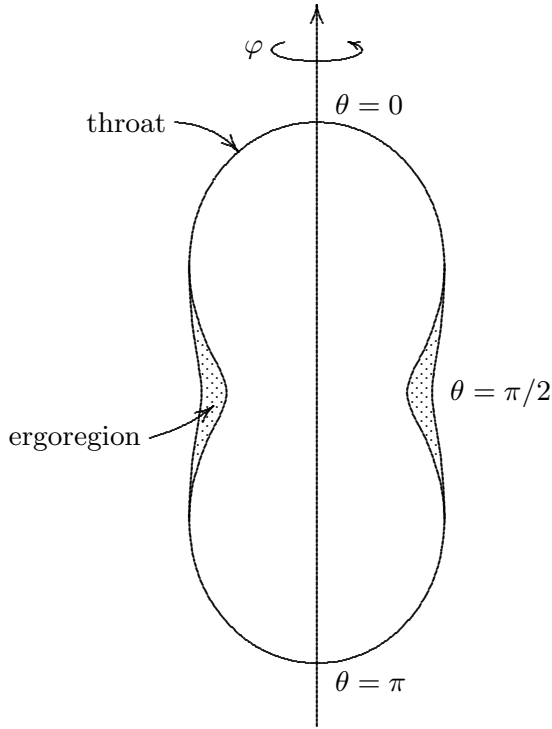
    \noindent Fig.~1.\enskip  Cross-sectional schematic of the wormhole 
    throat. The shaded region indicates the ergoregion, if it is present,
    surrounding the throat at the equator.}
  \vskip10mm
\egroup
\endinsert
%
%

If the rotation of the wormhole is sufficiently fast, $g_{tt}$ becomes 
positive in some region outside the throat, indicating the presence of 
an ergoregion where particles can no longer remain stationary with 
respect to infinity. For the above example, this occurs when 
$r^2=|2a\sin\theta|>1$, i.e., when $|a|>1/2$. Notice that the 
ergoregion does not completely surround the throat, but forms a `tube' 
around the equatorial region as illustrated in Fig.~1. This is 
characteristic of traversable wormholes: the ergoregion would 
necessarily intersect an event horizon at the poles, but since the 
latter is ruled out by definition, the ergoregion cannot extend to 
the poles. (It is also for this reason that the more traditional 
name of `ergosphere' is hardly appropriate here.)

When an ergoregion is present, it is possible to extract 
(rotational) energy from the wormhole by the Penrose process 
\refs{\PenroseI,\PenroseII}, which was originally proposed for the 
Kerr black hole. This process relies on the fact that time-like 
particles need not have positive energy in the ergoregion. Imagine an 
infalling particle breaking up into two inside the ergoregion. It is 
possible to arrange this breakup so that one of the resulting 
particles has negative total energy. The other particle can then 
travel out of the ergoregion along a geodesic, and would have more 
energy than what originally went in. However, the particle with 
negative energy would have an orbit confined entirely within the 
ergoregion. It is not possible for it to escape without gaining 
additional energy from some other source.

\newsec{Concluding remarks}

In this paper, I have presented the stationary, axisymmetric 
generalization of the Morris--Thorne wormhole. In particular,
I have written down the canonical form of the metric for such a
wormhole. This would allow one to describe rotating wormholes within 
a general framework. Although the null energy condition is 
generically violated at the throat, it is possible for geodesics 
falling through the wormhole to avoid this energy-condition-violating 
matter. Like rotating black holes, such wormholes can have 
ergoregions, from which energy can be extracted by the Penrose 
process.

Perhaps the most promising application of the results of this paper 
lies in the semi-classical or quantum regime. Unlike classical matter
fields, it is well-known that quantum fields {\it do\/} violate the 
energy conditions \EGJ. Indeed, self-consistent wormhole solutions 
have been found in Ref.~\HPS, in which the stress-energy tensor is 
that of a quantized scalar field \AHS. These wormholes have throats 
with radii of order of the Planck length, and could serve as a model 
for space-time foam. The possibility of generalizing this to wormholes 
with (slow) rotation has recently been discussed in Ref.~\Khatsym.

\listrefs
\bye